\documentclass[aps,prd,superscriptaddress,reprint,floatfix,onecolumn]{revtex4-2}

\usepackage{graphicx}

\usepackage{amsmath, amsthm, amssymb}
\usepackage{mathtools}
\usepackage{slashed}
\usepackage{bbold}
\usepackage[utf8]{inputenc}
\usepackage{empheq}
\usepackage{graphicx}
\usepackage{siunitx}
\usepackage{xcolor} 
\definecolor{KB}{rgb}{0, 0.5, 0}

\usepackage[draft]{changes}
\definechangesauthor[name={Natalia}, color=magenta]{nat}
\definechangesauthor[name={Konstantin}, color=KB]{kb}
\definechangesauthor[name={Konstantin}, color=teal]{kbNew}
\definechangesauthor[name={Igor}, color=orange]{iv}

\usepackage{tikz}
\usetikzlibrary{shapes.geometric, arrows,angles}

\usepackage[plainpages=false, colorlinks=true, anchorcolor=blue, linkcolor=blue, citecolor=blue, bookmarks=false]{hyperref}

\newcommand{\bra}[1]{\left\langle #1 \right\rvert}
\newcommand{\ket}[1]{\left\lvert #1 \right\rangle}

\newcommand{\defeq}{\vcentcolon =}
\newcommand{\rms}{r_\text{rms}}

\begin{document}
\title{Muon-Induced Nuclear Magnetic Moments in Spinless Muonic Atoms: A Simple Estimate}

\author{K. A. Beyer}
\email[Author to whom correspondence should be addressed: ]{konstantin.beyer@mpi-hd.mpg.de}
\affiliation{Max-Planck-Institut f\"ur Kernphysik, Saupfercheckweg 1, 69117 Heidelberg, Germany}

\author{N. S. Oreshkina}
\affiliation{Max-Planck-Institut f\"ur Kernphysik, Saupfercheckweg 1, 69117 Heidelberg, Germany}

\begin{abstract}
The magnetic field generated by a bound muon in heavy muonic atoms results in an induced nuclear magnetic dipole moment even for otherwise spinless nuclei. This dipole moment interacts with the muon, altering the binding energy of the muonic state. We investigate the relation of this simple, semi-classical-insriped approach to nuclear polarisation (NP) calculations. Motivated by the relative closeness of this simple estimate to evaluations of NP, we extract effective values for the nuclear magnetic polarisability, a quantity otherwise unknown, and put forward a simple back-of-the-envelope way to estimate the magnetic part of NP.
\end{abstract}
\maketitle

\section{Introduction}
Muonic atoms are of great interest because their orbitals, especially the low lying ones, are highly sensitive to nuclear structure~\cite{Wheeler1949,BorieRinker1982}. Spectroscopic measurements of $\mu-^{90}$Zr \cite{Phan:1985em}, $\mu-^{120}$Sn \cite{Piller:1990zza} and $\mu-^{208}$Pb \cite{Yamazaki:1979tf,Bergem:1988zz} reveal a striking, few-$\sigma$ discrepancy to theory predictions. The experimentally measured energy $E_a^\text{exp}$ of state $a$ does not match the prediction $E_a^\text{QED}$ based on quantum electrodynamics (QED). Especially the low lying muonic states $1s$, $2s$, and $2p$, clearly distinguishable from any electronic transitions because of their larger binding energies, pose a challenge.

Calculation of the nuclear polarisation (NP) contribution to the muonic energy levels is challenging and was therefore believed to be the cause for the discrepancy. Early studies were limited to the longitudinal component of NP, however, the inclusion of the transverse component did not alleviate the tension \cite{Tanaka:1992hto,Haga:2002zz}. Replacing the previous non-relativistic calculations with a relativistic mean field approach to the nucleus in Ref.~\cite{HAGA200571} further improved the understating of NP but did not serve to settle the problem. State-of-the-art NP calculations were performed in Ref.~\cite{Valuev:2022tau} taking into account dependence on nuclear models. Because of the persistence of the fine structure anomaly in these heavy muonic atoms, other QED effects have lately been scrutinised and re-evaluated. The latest one is the muonic self-energy in Ref.~\cite{Oreshkina:2022mrk}. Furthermore, efforts to explore Beyond the Standard Model contributions have not alleviated the tension either, see Ref.~\cite{Beyer:2023bwd}.

\begin{figure*}[t]
    \centering
    \includegraphics[width=0.95\textwidth]{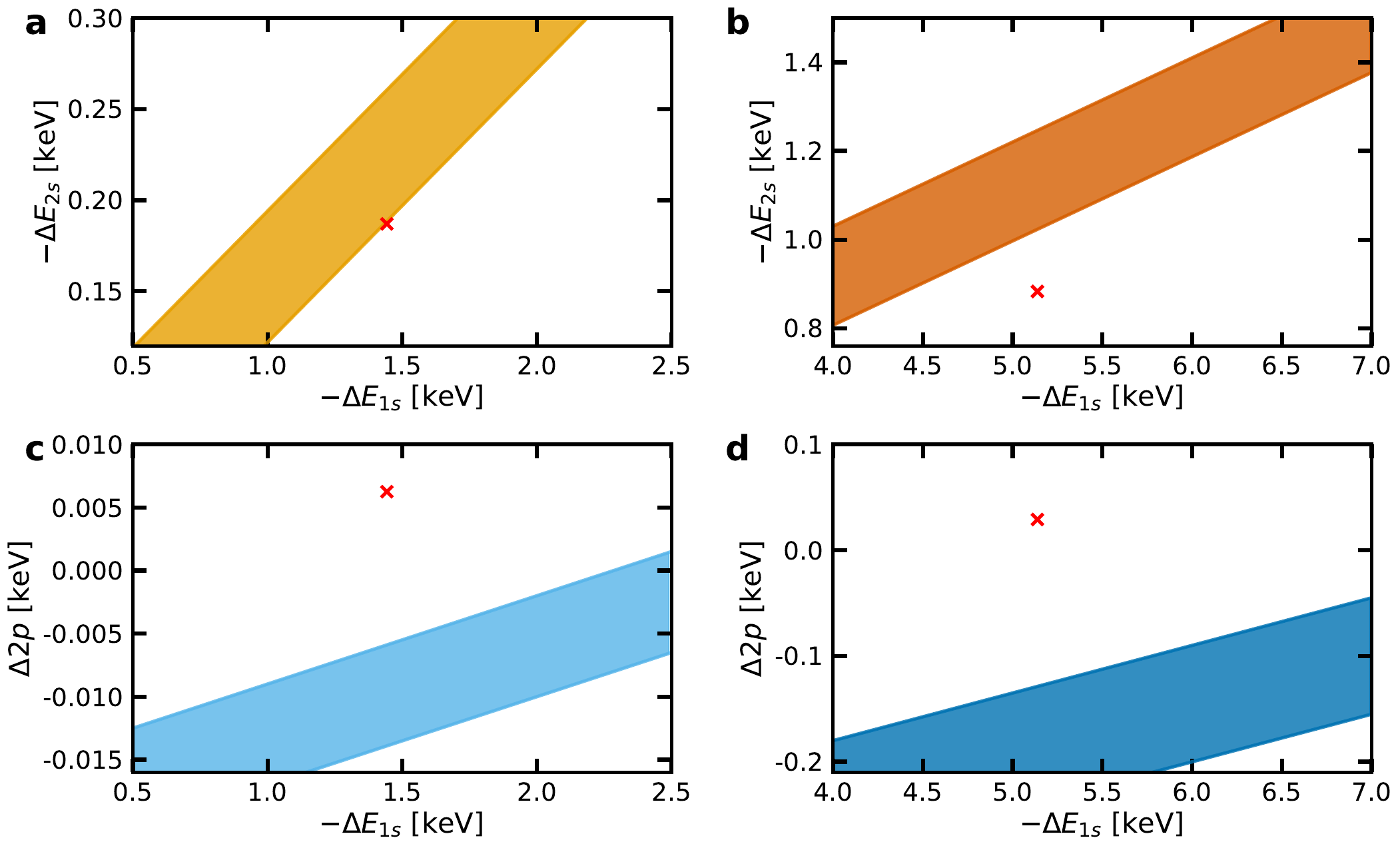}
    \caption{The coloured areas show the experimentally allowed range of shifts in $\Delta E_{2s}$ for $\mu-^{90}$Zr (a) and $\mu-^{208}$Pb (b) and $\Delta 2p$ for $\mu-^{90}$Zr (c) and $\mu-^{208}$Pb (d). The red crosses are the latest evaluations of the shifts from nuclear polarization in the SAMi parametrization, as found in Ref.~\cite{Valuev:2022tau} and self-energy corrections found in Ref.~\cite{Oreshkina:2022mrk}. The plots are adapted from Ref.~\cite{Phan:1985em} for $\mu-^{90}$Zr and Ref.~\cite{Yamazaki:1979tf,Bergem:1988zz} for $\mu-^{208}$Pb, respectively. The parameters as extracted are presented in table \ref{table:FigParam}.}
    \label{fig:ShiftsExp}
\end{figure*}

The puzzle concerns $\mu-^{90}$Zr, $\mu-^{120}$Sn and $\mu-^{208}$Pb, three nuclei with even number nucleons and vanishing nuclear angular momentum. These spherical nuclei do not possess a static magnetic moment and therefore no hyperfine and dynamical splitting which simplifies the interpretation of the spectroscopic measurements. Fig.~\ref{fig:ShiftsExp} shows the difference $\Delta a \equiv E_a^\text{QED} - E_a^\text{exp}$ {excluding NP} as coloured band with the thickness corresponding to the experimental uncertainty. The slope is a consequence of the dependence on the nuclear radius. The latest evaluations of contributions from nuclear polarisation (NP) \cite{Valuev:2022tau} and self-energy (SE) corrections \cite{Oreshkina:2022mrk} complete all leading order QED contributions. Being formerly unaccounted for in the QED prediction leading to the experimental region in Fig. \ref{fig:ShiftsExp}, they contribute to the difference and are shown as red crosses. It is evident that the anomaly is not resolved with the inclusion of these effects.

In the following we will motivate a simple estimate of the magnetic dipole contribution to the full NP correction based on a clear semi-classical picture. Because of its tight orbit, the bound muon induces a sizeable magnetic dipole moment in the nucleus. This results in a shift to the binding energy of the muon. We begin in \S~\ref{Sec:semi-classical} with a calculation of the induced magnetic dipole moment resulting from the bound muon's relativistic current and the binding energy shift. We then show in \S~\ref{Sec:IMM} how this contribution relates to NP and extract effective values for the nuclear magnetic dipole polarisability. Throughout the paper we work in natural units $\hbar=c=4\pi\varepsilon_0=1$ and $e<0$.

\begin{table*}[b]
\centering
\begin{tabular}{||c | c | c  | c  | c | r | r | r | r | r | r ||} 
 \hline
 Element &  $\mathcal{A}_{2p}^\text{Z}$ &  $\mathcal{A}_{2s}^\text{Z}$ & $\mathcal{B}_{2p,Z}^{(u,l)}$ [eV] & $\mathcal{B}_{2s,Z}^{(u,l)}$ [eV] & $\Delta E_{1s}^\text{NP}$[eV] & $\Delta E_{2s}^\text{NP}$[eV] & $\Delta 2p^\text{NP}$[eV] & $\Delta E_{1s}^\text{SE}$[eV] & $\Delta E_{2s}^\text{SE}$[eV] & $\Delta 2p^\text{SE}$[eV]\\ [0.5ex] 
 \hline 
 $^{90}$Zr &  $\SI{7e-3}{}$ & $0.15$ & $\SI{-16}{},\,\SI{-24}{}$ & $44,\,-28$ & $-1438$ & $-203$ & $5.7$ & $40$ & $16$ & $0.55$ \\[0.5ex] 
 $^{120}$Sn &  --- & --- & ---  & --- & $-2530$ & $-363$ & $18.7$ & $102$ & $-299$ & $-0.2$\phantom{0} \\[0.5ex]
 $^{208}$Pb &  $\SI{4.5e-2}{}$ & $0.19$ & $\SI{-360}{},\,\SI{-470}{}$ & $270,\,47$ & $-5727$ & $-1045$ & $59.1$ & $589$ & $162$ & $-31\phantom{.00}$ \\[0.5ex]
 \hline
\end{tabular}
\caption{Herein is presented a table displaying the parameters utilised for the nuclei discussed in the main text. The parameters for the compatibility region, parametrised as $\Delta  2p(\Delta E_{1s})=-\mathcal{A}_ {2p}  \Delta E_{1s} + \mathcal{B}_{2p}^{(u,l)}$ with $(u)$ [$(l)$] indicating the upper [lower] edge, were extracted from Ref.~\cite{Phan:1985em} ($\mu-^{90}$Zr), \cite{Piller:1990zza}($\mu-^{120}$Sn) and \cite{Yamazaki:1979tf,Bergem:1988zz} ($\mu-^{208}$Pb). The parametrisation for $2s$ is equivalent. The NP shifts are calculated in Ref.~\cite{Valuev:2022tau} and the SE corrections in Ref.~\cite{Oreshkina:2022mrk}. Note that the SE correction quoted here is the difference between the previous evaluation and the modern value.} 
\label{table:FigParam}
\end{table*}
\section{\label{Sec:semi-classical}semi-classical picture of muon-induced magnetic moment}
The wavefunction of a bound muon in the central Coulomb potential has a sizeable overlap with the nucleus. Heavier muons, compared to electrons, have a much smaller Bohr radius $r_\mu\propto (m_e/m_\mu) r_e$ and thus, their influence on the nucleus is less reduced by radial suppression. The orbiting muon produces a charged current $I\sim ev/2\pi r$ resulting in a magnetic dipole field $B\sim 2\pi I/r$. This field suffers from a $r^{-2}$ suppression which renders it negligible for electronic atoms. Muons in muonic atoms, having much tighter orbits, feel a magnetic field which is larger by a factor $(m_\mu/m_e)^2\sim \SI{4e4}{}$.

Under the assumption that the effect on the nucleus be small enough to treat it as a perturbation to the central Coulomb potential one can use the central-field muon wavefunctions known in the literature \cite{2006sham.book.....D}; the Dirac four spinor for state $a$ is
\begin{equation}
\label{Eq:MuonWavefunctions}
  \Psi_{ a }=\begin{pmatrix} g_a(r) \chi_{\kappa_a}^{m_a}(\theta,\varphi)\\if_a(r) \chi_{-\kappa_a}^{m_a}(\theta,\varphi) \end{pmatrix},
\end{equation}
with the bispinor
\begin{equation}
    \chi_\kappa^m(\theta,\varphi)=\begin{pmatrix} -\frac{\kappa}{\lvert\kappa\rvert}\sqrt{\frac{\kappa+\frac{1}{2}-m}{2\kappa+1}}\mathcal{Y}_{\lvert\kappa+\frac{1}{2}\rvert-\frac{1}{2}}^{ m-\frac{1}{2}}(\theta,\varphi)\\ \sqrt{\frac{\kappa+\frac{1}{2}+m}{2\kappa+1}}\mathcal{Y}_{\lvert\kappa+\frac{1}{2}\rvert-\frac{1}{2}}^{ m+\frac{1}{2}}(\theta,\varphi) \end{pmatrix}.
\end{equation}
Here, $\kappa_a\equiv(-1)^{j+l+1/2}\left(j+\frac{1}{2}\right)$ is the relativistic angular quantum number, $j$ is the total angular momentum quantum number with $m$ the projection onto the quantization axis, $l$ is the orbital momentum quantum number, and $\mathcal{Y}_a^b(\theta,\varphi)$ are normalized, complex spherical harmonics. Closed expressions for the radial components $g_a(r)$ and $f_a(r)$ are known in the limit of a pointlike nucleus, however, given the tight muon orbit, this simplification is not justified here. We will numerically solve for a spherical nucleus of constant density.

The charge density associated to a muon in state $a$ is
\begin{equation}
\label{Eq:ChargeDensity}
	\rho_a(\mathbf{r})=-e\Psi_a^\dagger(\mathbf{r})\Psi_a(\mathbf{r})=-e\left(g_a^2(r)\lvert\chi_\kappa^m(\theta,\varphi)\rvert^2 + f_a^2(r)\lvert\chi_{-\kappa}^m(\theta,\varphi)\rvert^2\right)
\end{equation}
and the current density
\begin{align}
\label{Eq:ChargeCurrent}
	 \mathbf{J}_a(\mathbf{r})&=-e\Psi_a^\dagger(\mathbf{r})\boldsymbol{\alpha}\Psi_a(\mathbf{r}) 
  =-ieg_a(r)f_a(r)\left((\chi_{\kappa}^{m}(\theta,\varphi))^\dagger \boldsymbol{\sigma}\chi^m_{-\kappa}(\theta,\varphi) -h.c\right),
\end{align}
with
\begin{equation}
    \boldsymbol{\alpha}=\gamma^0\boldsymbol{\gamma}=\begin{pmatrix}
        0&\boldsymbol{\sigma}\\\boldsymbol{\sigma}&0
    \end{pmatrix},
\end{equation}
$\gamma^\mu$ being Dirac matrices and $\boldsymbol{\sigma}$ -- Pauli matrices. The resulting static electric field is obtained as the solution to Maxwell's equations
\begin{align}
	\mathbf{E}_a(\mathbf{r})&=\int \frac{\mathbf{r}-\mathbf{r}'}{\lvert\mathbf{r}-\mathbf{r}'\rvert^3}\rho_a(\mathbf{r}')d^3\mathbf{r}'\equiv-e\bra{jm} \frac{\mathbf{r}-\mathbf{r}'}{\lvert\mathbf{r}-\mathbf{r}'\rvert^3}\ket{jm}.
\end{align}
The corresponding static magnetic field is
\begin{align}
\label{Eq:MagneticField}
	\mathbf{B}_a(\mathbf{r})&=-\int \frac{\mathbf{r}-\mathbf{r}'}{\lvert\mathbf{r}-\mathbf{r}'\rvert^3}\times\mathbf{J}_a(\mathbf{r}')d^3\mathbf{r}' = e \bra{jm}\frac{\mathbf{r}-\mathbf{r}'}{\lvert\mathbf{r}-\mathbf{r}'\rvert^3}\times\boldsymbol{\alpha}\ket{jm}.
\end{align}

As is expected for a symmetric charge distribution around the nucleus, the average electric field as seen by the nucleus vanishes, $\langle \mathbf{E}_a\rangle=0$, with 
\begin{equation} 
   \label{Eq:AverageField}
   \langle O\rangle\equiv\frac{\int_{V_\text{nucl}} O d^3\mathbf{r}}{V_\text{nucl}}=\frac{3}{4\pi R_0^3}\int_0^{R_0}dr r^2\int d^2\Omega   O,
\end{equation}
where we treat the nucleus as a homogeneously charged sphere of radius $R_0=\sqrt{5/3}\rms$ with $\rms$ the root-mean-square radii tabulated in Ref.~\cite{Schopper:2004qco}. This is to be expected of a symmetric charge distribution around the nucleus. In the following we will thus restrict ourselves to the magnetic field influence. The magnetic field of the muon will serve as a background field for the nucleus, inducing magnetic moments in the otherwise momentless magic nuclei in question. We are interested in the magnetic dipole moment and note that, by definition, the ratio between magnetic dipole moment $\mathbf{M}_a$ and magnetic field $\mathbf{B}_a$ is the magnetic dipole polarisability $\beta$. Thus, the induced magnetic dipole moment can be expressed as
\begin{equation}
\label{Eq:DefMa}
    \mathbf{M}_a\equiv\beta \mathbf{B}_a.
\end{equation}

This induced magnetic dipole moment leads to a hyperfine shift in the energy levels of the muon through the interaction
\begin{equation}
\label{Eq:MagnMomCoupling}
    \mathcal{H}=e\boldsymbol{\alpha}\cdot\mathbf{A}=-e\frac{\mathbf{M}\cdot(\mathbf{r}\times\boldsymbol{\alpha})}{r^3}.
\end{equation}
Together with Eq.~\eqref{Eq:DefMa} we find an energy shift
\begin{equation}
    \Delta E_a^M\equiv \bra{jm}\mathcal{H}\ket{jm} =-e^2 \beta \langle\bra{jm}\frac{\mathbf{r}-\mathbf{r}'}{\lvert\mathbf{r}-\mathbf{r}'\rvert^3}\times\boldsymbol{\alpha}\ket{jm}\rangle\cdot\bra{jm} \frac{\mathbf{r}\times\boldsymbol{\alpha}}{r^3} \ket{jm}.
\end{equation}
In the limit of a pointlike nucleus the energy shift reduces to
\begin{equation}
\label{Eq:semi-classicalErgShift}
    \Delta E_a^M =-e^2 \beta \left | \bra{jm} \frac{\mathbf{r}\times\boldsymbol{\alpha}}{r^3} \ket{jm} \right |^2.
\end{equation}

We will show in the following, that this expression corresponds to a piece of the magnetic dipole moment contribution to NP.

\section{\label{Sec:IMM}Induced Magnetic Moment Part of Nuclear Polarisation}
The above semi-classical calculation gives an intuitive picture of the interaction taking place between the nucleus and muon. The induced nuclear magnetic moment contribution is part of the general second-order electromagnetic interaction between the nucleus and muon which is known as nuclear polarisation. We would like to show the correspondence between the two expressions explicitly here.

The electromagnetic interaction can be divided into a charge-density and current-density interaction
\begin{equation}
V\equiv\gamma^0 \gamma_\nu A^\nu = A_0 - \boldsymbol{\alpha}\cdot\mathbf{A} .
\end{equation}
The system is a direct product of the bound muon and the nucleons $\ket{aA}=\ket{a}\ket{A}$, and we note that for us the initial and final nuclear state is the ground state $\ket{A}=\ket{0}$. This can be argued based on the fact that the energies involved are not high enough to cause excitation of all nuclei in the target and the aforementioned experiments did not see a forest of transitions as would be expected for a mixture of ground and excited nuclei. In the following we will focus on the magnetic dipole moment contribution, which sits in the current piece of this interaction. An expansion in multipole moments leaves \cite{PhysRev.97.380}
\begin{equation}
	V_M\defeq e\sum_{j=1}^Z \boldsymbol{\alpha}\cdot\mathbf{A}(\mathbf{r},\mathbf{r}_j) = -ie\sum_{j=1}^Z \boldsymbol{\alpha}\cdot \left( \sum_{l=0}^\infty\sum_{m=-l}^l \sqrt{\frac{4\pi}{2l+1}}\frac{(\hat{\mathbf{L}}\mathcal{Y}_l^m(\theta,\phi))}{lr^{l+1}}M^{(l)}_{-m}(\mathbf{r}_j)\right).
\end{equation}
Here,  $\hat{\mathbf{L}}\equiv -i\mathbf{r}\times\boldsymbol{\nabla}$ is the angular momentum operator and the multipole operators are
\begin{equation}
	M^{(l)}_{m}(\mathbf{r}_j)\equiv\mu_N\sqrt{\frac{4\pi}{2l+1}}\left(\boldsymbol{\nabla} r_j^l  \mathcal{Y}_l^m(\theta_j,\phi_j)\right)\cdot\left(g_l\frac{2}{l+1}\hat{\mathbf{L}}_j+g_S \hat{\mathbf{S}}_j\right).
\end{equation}
$\hat{\mathbf{S}}_j$ is the spin operator of the nucleon $j$, and $\mathcal{Y}_l^m(\theta,\phi)$ are complex spherical harmonics.

Being interested in the magnetic dipole moment, we restrict the multipole expansion to $l=1$
\begin{equation}
\label{Eq:DipolePotential}
	V_M^{(l=1)}= -ie\sum_{j=1}^Z \boldsymbol{\alpha}\cdot \left(\sum_{m=-1}^1 \sqrt{\frac{4\pi}{3}}\frac{\left(\hat{\mathbf{L}}\mathcal{Y}_1^m(\theta,\phi)\right)}{r^{2}} M^{(1)}_{-m}(\mathbf{r}_j)\right).
\end{equation}
The second order contribution to the muon's binding energy is
\begin{equation}
	\Delta E_M^{(l=1)}=\sum_{nN}\frac{\bra{a0}V_M^{(l=1)}\ket{Nn}\bra{nN}V_M^{(l=1)}\ket{0a}}{E_{a0}-E_{nN}}.
\end{equation}
The matrix element can be separated into a nuclear part consisting of the multipole operators $M^{(l)}_{m}(\mathbf{r}_j)$, and a muonic part containing the rest of the expression. We define the magnetic dipole polarisability in analogy to the electric polarisability \cite{PhysRevResearch.3.L032032}, 
\begin{equation}
	\beta^{(l=1)}_{m_1 m_2}\defeq \sum_{j=1}^Z\sum_{i=1}^Z \sum_N \frac{\bra{0}M^{(1)}_{-m_1}(\mathbf{r}_j)\ket{N}\bra{N}M^{(1)}_{-m_2}(\mathbf{r}_i)\ket{0}}{E_N-E_0},
\end{equation}
where we chose $E_N-E_0$, leaving the polarisability strictly positive. Note that for the nuclei in question $\bra{0}M^{(1)}_{-m_1}(\mathbf{r}_j)\ket{0}=0$ and therefore we only sum all $N\neq 0$. This allows us to make the following expansion 
\begin{equation}
    \Delta E_M^{(l=1)}=\sum_{j=1}^Z\sum_{i=1}^Z \sum_N\frac{\bra{0}M^{(1)}_{-m_1}(\mathbf{r}_j)\ket{N}\bra{N}M^{(1)}_{-m_2}(\mathbf{r}_i)\ket{0}}{E_N-E_0}\bra{a}\hat{O}\ket{a}\bra{a}\hat{O}\ket{a}\left( 1+\sum_{n\neq a}\frac{\bra{a}\hat{O}\ket{n}\bra{n}\hat{O}\ket{a}}{\bra{a}\hat{O}\ket{a}\bra{a}\hat{O}\ket{a}}\frac{1}{1+\frac{E_a-E_n}{E_0-E_N}}\right)
\end{equation}
where $\hat{O}$ is the muonic part of the operator in eq.~\eqref{Eq:DipolePotential}. We note that for $a = 1s$, $2p_{1/2}$, $2p_{3/2}$, $3p_{1/2}$, and $3p_{3/2}$, the first term ($n=a$) of the sum over the bound part of the muonic spectrum dominates over the next biggest term which is smaller by at least a factor $1/2$. We therefore restrict ourselves to $n=a$ such that we can write
\begin{equation}
\label{Eq:SecondOrderMagenticErgShift}
	\Delta E_M^{(l=1)}=\frac{4\pi}{3}e^2\sum_{m_1=-1}^1\sum_{m_2=-1}^1 \beta^{(l=1)}_{m_1 m_2}\bra{a} \boldsymbol{\alpha}\cdot \left(\frac{\hat{\mathbf{L}}}{r^{2}}\mathcal{Y}_1^{m_1}(\theta,\phi)\right) \ket{a}\bra{a} \boldsymbol{\alpha}\cdot \left(\frac{\hat{\mathbf{L}}}{r^{2}}\mathcal{Y}_1^{m_2}(\theta,\phi) \right)\ket{a}+\mathcal{O}(\varepsilon).
\end{equation}

It can easily be shown that $$\frac{\Delta E_M^{(l=1)}}{\Delta E_a^M}=4\frac{\beta^{(l=1)}_{00}}{\beta}.$$ 
Thus, the aforementioned simple, semi-classical contribution corresponds to the first term in the sum over the muonic spectrum of the NP correction.  Performing the full sum over the spectrum is challenging, especially the continuum contributions are a serious obstacle. We would like to ask the opposite question and understand how far the above simple picture can take us. in order to achieve this we aim to compare the simple picture with results obtained from much more sophisticated numerical approaches as in \cite{Valuev:2022tau}.

We begin by relaxing the above assumption $n=a$ to $E_n=E_a$. This preserves the interpretation along the semi-classical picture of section \ref{Sec:semi-classical} but takes into account the projection quantum number $m$ part of the sum over intermediate states, $\sum_n \Rightarrow \sum_{m'}$ and Eq. \eqref{Eq:semi-classicalErgShift} reduces to
\begin{equation}
	\Delta E_a^M =-e^2 \beta \sum_{m'} \bra{jm}\frac{\mathbf{r}\times\boldsymbol{\alpha}}{r^3}\ket{jm'}\cdot\bra{jm'} \frac{\mathbf{r}\times\boldsymbol{\alpha}}{r^3} \ket{jm}.
\end{equation}
Using the Wiegner-Eckard theorem we may simplify
\begin{equation}
\label{Eq:ErgShift}
	\Delta E_a^M =-e^2 \beta \frac{\kappa_a^2}{j(j+1)}(\mathcal{I}_a^{M})^2,
\end{equation}
with
\begin{equation}
	\mathcal{I}_a^{M} \defeq \int g_a(r) f_a(r) dr.
\end{equation}
To account for the finite nuclear size we replace
\begin{equation}
\label{Eq:FinSizeIntegral}
	\mathcal{I}_a^{M} = \int g_a(r) f_a(r) dr \quad \Rightarrow \quad \mathcal{I}_a^{M}\defeq \int_0^{R_0} dr f_a(r) g_a(r) \left(\frac{r}{R_0}\right)^3 + \int_{R_0}^{\infty} dr f_a(r) g_a(r),
\end{equation}
see table \ref{table:Integrals} for numerical evaluation of the integrals.

\begin{table*}[t]
\centering
\begin{tabular}{c || c | c | c  | c  | c |  c  |  c  |  c  |  c  | c } 
 State & $1s$ [Pb] & $2p_{1/2}$ [Pb] &$2p_{3/2}$ [Pb]& $2s$ [Pb] & $3p_{1/2}$ [Pb]&$3p_{3/2}$ [Pb]&$1s$ [Zr] & $2p_{1/2}$ [Zr] &$2p_{3/2}$ [Zr] & $2s$ [Zr]\\\hline
   $\mathcal{I}_{a}^M$ [MeV$^2$] & $-44.5$ &$ 23.8 $&$-26.8 $& $-5.5 $&$ 5.1 $&$ -7.2 $&$-36.3 $&$ 8.7 $&$ -5.4 $&$ -4.8 $\\
  $\Delta E_a^M$ [eV] & $2.4$ &$ 0.7 $&$0.7 $& $0.04 $&$ 0.03 $&$ 0.05 $&$1.6 $&$ 0.1 $&$ 0.03 $&$ 0.03 $\\
  $\beta_a^\text{eff}$ [fm$^3$] & $5.2-7.6$ &$ 4.4-7.2$&$2.4-4.8$& $-43.1$&$ 41.1 $&$ 12.3 $&$2.8-3.2$&$1.2-2.0$&$3.3$&$19.9$
\end{tabular}
\caption{Numerical evaluation of the remaining radial integral Eq. \eqref{Eq:FinSizeIntegral} for spherical, homogeneous nuclear charge distribution. The resulting energy shift Eq. \eqref{Eq:ErgShift} is tabulated for $\beta = 1\text{fm}^3$. The third row shows the effective nuclear magnetic polarisability as defined in Eq. \eqref{Eq:EffBeta}.} 
\label{table:Integrals}
\end{table*}

The final energy shifts are tabulated in table \ref{table:Integrals} for $\beta = 1\text{fm}^3$; Scaling the results to any $\beta$ is trivial because of the linear dependence of the energy shift on the polarisability. Unfortunately, the nuclear magnetic polarisability is not known for most nuclei, however, a reasonable estimate can be obtained by comparison with the electric polarisability \cite{Gorchtein:2015eoa}. 

When comparing with ref. \cite{Valuev:2022tau}, in which the NP correction $\Delta E_a$ was evaluated under consideration of the full muon spectrum, we find that our simple estimate is surprisingly close. We can for example take the result of ref. \cite{Valuev:2022tau} and compare it with our simple estimate to get an effective polarisability
\begin{equation}
\label{Eq:EffBeta}
    \beta_a^\text{eff} = \Delta E_a\frac{j(j+1)}{e^2\kappa_a^2 (\mathcal{I}_a^M)^2},
\end{equation}
which is tabulated in the last row of table \ref{table:Integrals}.
We see that the first three states $1s$, $2p_{1/2}$, and $2p_{3/2}$ are consistent with $\beta_a^\text{eff}$ being approximately constant. This trend breaks once the next higher states $2s$, $3p_{1/2}$, and $3p_{3/2}$ are considered. We attribute this to the breakdown of the single level approximation for these states. The higher lying states get contributions from cross terms with the lower lying states and states of similar binding energy as them. E.g. $3p_{3/2}$ will have a contribution from $n=3p_{3/2}$, $3p_{1/2}$, $2p_{3/2}$, and $2p_{1/2}$. Even if each individual contribution is small, accumulated they may contribute significantly.

\section{Conclusion}
We have shown that the magnetic dipole part of the complicated NP correction to muonic atoms is surprisingly well approximated by a simple semi-classical interaction. The magnetic field of the orbiting muon induces a magnetic dipole moment into the otherwise spinless nucleus. In the field of this induced magnetic moment, the muon's binding energy is slightly altered. Comparison with full evaluations, taking into account the full muon spectrum, we find the values to be surprisingly close. We take this as motivation to extract effective values for the nuclear magnetic polarisability, a quantity otherwise unknown.

\section*{Acknowledgements}
We thank Igor A. Valuev, Mikhail Gorchtein, Vladimir A. Yerokhin, Vladimir A. Zaytsev, and Matteo Tamburini for fruitful discussions.

\bibliography{Bibliography}
\appendix
\widetext

\end{document}